\documentclass[conference]{IEEEtran}
\IEEEoverridecommandlockouts
\usepackage{amsmath,amssymb,amsfonts}
\usepackage{algorithmic}
\usepackage{hyperref}
\usepackage{graphicx}
\usepackage{textcomp}
\usepackage{xcolor}
\usepackage{todonotes}
\usepackage[shortlabels]{enumitem}
\usepackage{flushend}
\def\BibTeX{{\rm B\kern-.05em{\sc i\kern-.025em b}\kern-.08em
    T\kern-.1667em\lower.7ex\hbox{E}\kern-.125emX}}
\usepackage[style=ieee, citestyle=numeric-comp, mincitenames=1, maxcitenames=1, maxnames=3]{biblatex}
\bibliography{literature}

\begin{document}

\title{Vulnerability Handling of AI-Generated Code -- Existing Solutions and Open Challenges
\thanks{
This work has been funded by the Deutsche Forschungsgemeinschaft (DFG, German Research Foundation) – Project-ID 528745080 - FIP 68. The authors alone are responsible for the content of the paper.}
}
\author{
	\IEEEauthorblockN{Sabrina Kaniewski, Dieter Holstein, Fabian Schmidt, Tobias Heer} 
	\IEEEauthorblockA{
		Esslingen University of Applied Sciences, Germany, \\ 
        \textit{\{sabrina.kaniewski, dihoit00, fabian.schmidt, tobias.heer\}@hs-esslingen.de}
	}
}
\maketitle

\iftrue 

\newlength{\bottommargin}
\bottommargin=\paperheight
\addtolength{\bottommargin}{-1in}
\addtolength{\bottommargin}{-\voffset}
\addtolength{\bottommargin}{-\topmargin}
\addtolength{\bottommargin}{-\headheight}
\addtolength{\bottommargin}{-\headsep}
\addtolength{\bottommargin}{-\textheight}
\begin{tikzpicture}[remember picture,overlay,shift={(current page.south)}]\node[draw,red,rectangle,thick,font=\small] at (0,.5\bottommargin) {
		\begin{minipage}{\textwidth}
Accepted for publication in Proceedings of \emph{IEEE Artificial Intelligence x Science, Engineering, and Technology (AIxSET)}, \\ Laguna Hills, California, USA, September 30 - October 2, 2024 \\
			\textbf{\copyright 
2024 IEEE.  Personal use of this material is permitted.  Permission from IEEE must be obtained for all other uses, in any current or future media, including reprinting/republishing this material for advertising or promotional purposes, creating new collective works, for resale or redistribution to servers or lists, or reuse of any copyrighted component of this work in other works.} 
		\end{minipage}};
\end{tikzpicture}%
\fi

\vspace{-1em}

\begin{abstract}
The increasing use of generative Artificial Intelligence (AI) in modern software engineering, particularly Large Language Models (LLMs) for code generation, has transformed professional software development by boosting productivity and automating development processes. This adoption, however, has highlighted a significant issue: the introduction of security vulnerabilities into the code. These vulnerabilities result, e.g., from flaws in the training data that propagate into the generated code, creating challenges in disclosing them. Traditional vulnerability handling processes often involve extensive manual review. Applying such traditional processes to AI-generated code is challenging. AI-generated code may include several vulnerabilities, possibly in slightly different forms as developers might not build on already implemented code but prompt similar tasks.
In this work, we explore the current state of LLM-based approaches for vulnerability handling, focusing on approaches for vulnerability detection, localization, and repair. We provide an overview of recent progress in this area and highlight open challenges that must be addressed in order to establish a reliable and scalable vulnerability handling process of AI-generated code.
\end{abstract}

\section{Introduction}
\begin{figure*}[!t]
        \centering  
        \includegraphics[width=1\linewidth]{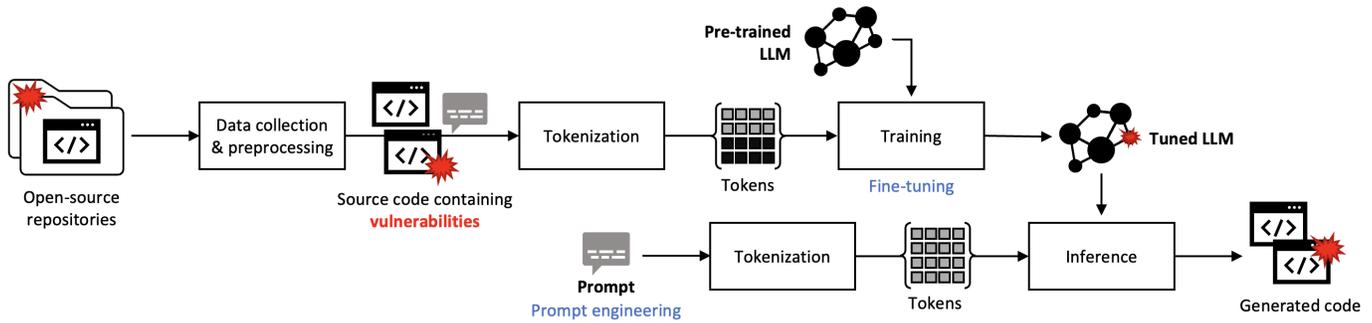}
        \vspace{-1.5em}
        \caption{Training and inference pipeline for code generation with LLMs. Vulnerabilities in training data transfer to generated code.}
        \label{fig:motivation}
        \vspace{-0.75em}
\end{figure*}

The use of AI in modern software engineering has led to a significant transformation in the industry, especially with the adoption of generative AI for tasks such as code generation. 
Particularly prominent are LLMs, e.g., ChatGPT~\cite{chatgpt} and Codex~\cite{chen2021codex}, that generate code snippets based on given functionality prompts, promising to enhance productivity.
The adoption of AI-based tools for code generation, however, has brought to light the prevalence of security vulnerabilities in AI-generated code~\cite{pearce2022asleep, liu2024no}. 
These vulnerabilities stem, e.g., from vulnerabilities in the training data, as many models are trained on faulty code from public repositories~\cite{pearce2022asleep}.
Consequently, these vulnerabilities can transfer into the generated code during inference, as visualized in Figure~\ref{fig:motivation}.
LLMs also have been found to generate hallucinations~\cite{liu2024exploring} and code smells~\cite{siddiq2022empirical}, such as redundancy and inconsistency, leading to security vulnerabilities.
In large software projects, redundancy becomes more evident.
Developers may prompt similar tasks without considering pre-existing implementations, scattering semantically identical but syntactically different vulnerabilities across a vast codebase, making them harder to remediate. 

\begin{figure}[!b]
        \centering  
        \vspace{-1em}
        \includegraphics[width=0.8\linewidth]{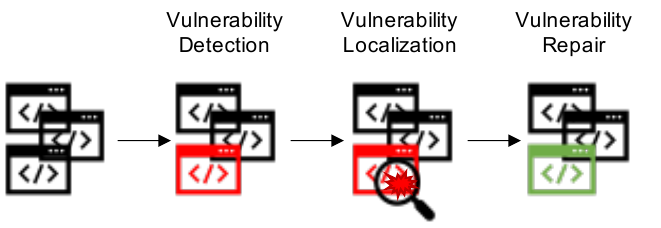}
        \vspace{-0.75em}
        \caption{The process of handling security vulnerabilities in software.}
        \label{fig:pip}
\end{figure}

The process of handling vulnerabilities and maintaining code security and quality is well formalized in the life cycle of software. 
Traditionally, vulnerability handling has been a manual process driven, e.g., by Common Vulnerabilities and Exposures (CVE) descriptions.
Once a vulnerability is reported, the process works through the steps of vulnerability detection, localization, and repair, see Figure \ref{fig:pip}.
Vulnerability detection determines whether a product is affected by known vulnerabilities, e.g., because it uses a vulnerable library, requiring extensive code review. 
Vulnerability localization pinpoints the lines of code containing these vulnerabilities, using methods such as static code analysis. 
Finally, vulnerability repair involves fixing the identified vulnerabilities with suitable patches while preserving the integrity and functionality of the code, e.g., replacing the vulnerable library with a fixed version.
When using AI-based tools for code generation, the handling of vulnerabilities becomes more complex due to redundant code and poor referencing.
It is not sufficient for developers to merely update a library or component; rather, they must conduct exhaustive code analysis to identify all affected areas where similar generated code is used.
Consequently, developers must invest considerable manual effort or resort to AI-based solutions to address this challenge more efficiently. 

In this work, we explore the current state of AI use for the handling of security vulnerabilities, focusing on recent LLM-based approaches proposed for vulnerability detection, localization, and repair.
The goal of this work is to provide a concise overview of current research approaches in the field. We further aim to highlight open challenges that must be addressed to establish a vulnerability handling process that is concerned with the use of AI in software engineering.



\section{LLMs for Vulnerability Handling}
\label{sec:llms}
Researchers primarily use fine-tuning or prompt engineering, as highlighted in Figure~\ref{fig:motivation}, to tailor LLMs to the requirements of vulnerability handling tasks.
In the following, we present recent LLM-based approaches for vulnerability detection, localization, and repair, including preliminary research, outline recent progress and remaining limitations.


\subsection{Vulnerability Detection}
Vulnerability detection involves identifying and classifying the presence or type of vulnerabilities in the given code.
Steenhoek et al.~\cite{steenhoek2024comprehensive} evaluate the capabilities of multiple LLMs, including GPT-4~\cite{achiam2023gpt4}, 
WizardCoder~\cite{luo2023wizardcoder} and CodeLlama~\cite{roziere2023codellama}, for vulnerability detection using in-context learning (ICL)~\cite{brown2020gpt3} and chain-of-thought (CoT) prompting~\cite{wei2022chain}. 
ICL guides responses with input-output examples, while CoT prompting uses step-by-step reasoning with contextual information.
The authors identify limitations in distinguishing between non-vulnerable and vulnerable code, with some models close to random guessing, and handling of longer code due to input limitations.
Similarly, Fu et al.~\cite{fu2023chatgpt} evaluate GPT-4, among other models, for classifying vulnerability types using different prompting methods. 
They find that the performance lags behind specialized, fine-tuned models, such as CodeBERT~\cite{feng2020codebert}. 
Zhou et al.~\cite{zhou2024large}, in contrast, report enhanced capabilities of GPT-4 compared to CodeBERT when incorporating external knowledge from different vulnerability types into CoT prompts. 
Zhang et al.~\cite{zhang2024prompt} and Nong et al.~\cite{nong2024chain} also observe improved performance 
with CoT prompts incorporating structural and sequential information or vulnerability-semantics. 
{GPT-4}, however, struggles with detecting complex vulnerabilities.
Furthermore, the performance improves only when applied to synthetic datasets, not complex real-world datasets.
Ullah et al.~\cite{ullah2024llms} use GPT-4 and PaLM2~\cite{anil2023palm2} to reason about detected vulnerabilities, observing non-deterministic responses and a lack of contextual reasoning. 
Shestov et al.~\cite{shestov2024finetuning} demonstrate the effectiveness of fine-tuning pre-trained LLMs, particularly WizardCoder, for vulnerability detection. 
Ding et al.~\cite{ding2024vulnerability} address that, 
while fine-tuning and ICL~improve detection accuracy, results suggest that models rely on superficial text patterns rather than understanding underlying security implications, which hinders their use in real-world code scenarios.
Risse and Böhme~\cite{risse2023limits} evaluate model robustness against code transformations, e.g., renaming variables. They find that models overfit to specific transformations used in training data, failing to generalize to new settings, which is crucial for a safe integration into software engineering processes.


\subsection{Vulnerability Localization}
The objective of vulnerability or fault localization is to accurately identify the lines of code most likely to contain vulnerabilities. 
Wu et al.~\cite{wu2023large} evaluate LLMs for vulnerability localization, finding that GPT-4 outperforms traditional static and dynamic code analysis methods with prompts incorporating function-level context and error logs. However, performance degrades with larger code contexts.
Similarly, Zhang et al.~\cite{zhang2024empirical} evaluate, e.g., GPT-4, CodeLlama, WizardCoder, CodeGen~\cite{nijkamp2022codegen}, CodeBERT, and GraphCodeBERT~\cite{guo2020graphcodebert}, using different prompt designs and fine-tuning.
While outperforming traditional tools, they encounter limitations with context lengths and less common vulnerability types.
Qin et al.~\cite{qin2024agentfl} utilize a multi-agent system based on ChatGPT~\cite{chatgpt} with specialized agents for different tasks for project-level fault localization. While outperforming other LLM-based approaches, difficulties with larger code projects remain.
Widyasari et al.~\cite{widyasari2024demystifying} use ChatGPT with CoT prompting to generate fault explanations. Their method increases successful fault localization and generates mostly correct explanations, though quality improvements are needed.
Yang et al.~\cite{yang2024large} enhance code understanding of LLMs through bidirectional adapter layers to localize faults without test cases. By fine-tuning CodeGen models, they significantly outperform existing machine-learning-based techniques but may still struggle with complex contextual information due to reliance on pre-training.


\subsection{Vulnerability Repair}
Vulnerability repair represents a sequence-to-sequence task, i.e., taking a sequence of code containing a vulnerability and generating one where the vulnerability is patched.
Fu et al.~\cite{fu2023chatgpt} compare GPT-4 to baselines such as CodeBERT and GraphCodeBERT for generating patches using ICL. 
Similarly, Le et al.~\cite{le2024study} evaluate ChatGPT and Bard \cite{bard} with context-enriched prompts.
Structured prompts enhance the model's generation of effective repair patches, though performance varies by context and vulnerability type and lags behind fine-tuned models.
Pearce et al.~\cite{pearce2023examining} and Nong et al.~\cite{nong2024chain} evaluate several LLMs,  
using enhanced prompt designs. 
The models generate patches with improved quality in controlled scenarios but struggle with complex real-world vulnerabilities. 
Alrashedy and Aljasser~\cite{alrashedy2024can} prompt GPT-4 and CodeLlama to iteratively generate and refine patches based on feedback given by a static code analysis tool. 
While the approach shows improved results, it faces limitations in identifying all vulnerabilities and preserving functionality. 
Zhang et al.~\cite{zhang2023pre} assess pre-trained models 
against traditional techniques, finding challenges with complex functions.
While fine-tuning improves performance, training data size and quality issues persist.

\vspace{1em}
\section{Open Challenges}
\label{sec:open}
As AI-based code generation tools introduce security vulnerabilities into the codebase in a unique way, i.e., in redundant
form and with poor referencing, there is a critical need for processes that can effectively remediate them.
Integrating LLMs into the handling process faces several open challenges that persist across the studies introduced above.

To effectively utilize LLMs for vulnerability handling, all context relevant to the detection, localization, or repair task has to be provided to the LLM.
Providing the code to be analyzed is already challenging for larger codebases and project-level repositories, i.a., due to the token limitations of some models \cite{nong2024chain,steenhoek2024comprehensive}. 
To provide additional information for the task, e.g., vulnerability descriptions or test case outputs, efficient prompt designs are required.
The vast number of vulnerability descriptions and the rapid emergence of new, zero-day vulnerabilities, however, present ongoing challenges. 
In recent years, over $20,000$ new CVEs have been published each year~\cite{cve_metrics}.
As a result, it is practically infeasible to~include all CVE descriptions within the model's prompt.
Hence, research needs to investigate strategies for adapting LLMs to large inputs and vast amounts of vulnerability descriptions in a structured manner and within a reasonable time to keep pace with the evolving landscape of software vulnerabilities.
One promising approach that requires further investigation is Retrieval-Augmented Generation (RAG)~\cite{lewis2020retrieval}. 
RAG enhances generative models by incorporating relevant external information into the prompt. For example, RAG can be used to retrieve CVE descriptions with high similarity to the code to be analyzed from an external CVE database, see similar works~\cite{du2024vul,zhou2024out}.

Training and fine-tuning are essential to equip LLMs with the domain-specific expertise required for effective vulnerability handling \cite{alrashedy2024can, ding2024vulnerability}. 
However, training LLMs is very time-consuming and can extend up to several weeks~\cite{xu2022polycoder}, highly dependent on the model size, the computational resources available, and the amount of training data.
The process is also costly, not only in hardware and energy consumption but resources required to curate diverse and high-quality vulnerability datasets.
While efforts have been made to train or fine-tune LLMs on CVE datasets, e.g.,~\cite{bhandari2021cvefixes,ding2024vulnerability},
these datasets do not cover the entirety of reported vulnerabilities, variations of human-made and generated code, complex vulnerabilities, and less frequent CVE types. 
This required diversity ensures that the models can generalize across different scenarios and recognize subtle patterns in the data. Synthetic and hand-crafted datasets used in most studies, limited to a selection of CVEs, fall short of emulating real-world scenarios, often involving nuanced and multifaceted vulnerabilities.
Thus, developing efficient data collection and pre-processing methods is crucial.
By leveraging automated data collection and advanced pre-processing techniques, researchers can curate comprehensive datasets that better represent the diversity and complexity of real-world vulnerabilities, ultimately improving the models' capabilities in practical applications.

LLMs rely on pattern recognition, which limits their ability to understand security-related concepts \cite{ullah2024llms} and to accurately identify, localize, and repair vulnerabilities. 
This limitation is further evident in their non-deterministic outputs when prompted multiple times \cite{ullah2024llms,nong2024chain}.
Overfitting to specific patterns during training can also restrict the models' applicability and generalization to previously unseen and real-world vulnerabilities \cite{risse2023limits}. 
To ensure a reliable vulnerability handling process, it is essential to enhance the models' understanding of security-related concepts, aiming to avoid generating code with known vulnerabilities in the first place.
Achieving this requires a long-term strategy that includes refining training data, implementing robust monitoring systems, and adopting continual learning approaches~\cite{shi2024continual}. 
Continual learning, in particular, can help models adapt to evolving security practices by incrementally updating their knowledge base, thereby improving their effectiveness and reliability. 

\vspace{1em}
\section{Conclusion}
\label{sec:conclusion}


Software engineering is witnessing an increasing incorporation of the use of AI, particularly LLMs for code generation.
However, semantically identical but syntactically different vulnerabilities in AI-generated code make it challenging to address all vulnerabilities effectively with traditional methods.
Despite advancements in utilizing LLMs to support vulnerability handling, there is currently no comprehensive~process capable of managing the scale and complexity of reported vulnerabilities.
Existing research addresses only partial aspects of the problem, leaving the handling of vulnerabilities limited and incomplete.
Therefore, a more holistic and adaptable approach is needed to establish a reliable and scalable vulnerability handling process. 
This approach must include the continuous refinement of techniques and data for the handling and mitigation of vulnerabilities in both AI-generated and traditional codebases.
By addressing the open challenges presented, we can take a step towards maintaining secure and high-quality code in the evolving landscape of software engineering.

\vspace{1em}

\printbibliography

\end{document}